\RequirePackage{fix-cm}
\documentclass[natbib, smallcondensed]{svjour3}     
\smartqed  
\usepackage{graphicx}
\usepackage{subcaption}
\usepackage{float}
\usepackage{rotating}
\usepackage{lscape}
\usepackage{capt-of}
\usepackage{multirow}
\begin{document}

\title{Empirical Characterization of Graph Sampling Algorithms
}


\author{Muhammad Irfan Yousuf, Izza Anwer, Raheel Anwar}

\authorrunning{M. I. Yousuf et al.}

\institute{M. I. Yousuf \at
              Department of Computer Science (New Campus) \\
              University of Engineering and Technology, Lahore, Pakistan\\
              Tel: +92-42-37951901\\
              \email{irfan.yousuf@uet.edu.pk}\\           
              I. Anwer \at
              Department of Transportation Engineering and Management\\
              University of Engineering and Technology, Lahore, Pakistan\\
              R. Anwar \at
              Karl Franzens Universität, Graz, Austria
}

\date{Received: date / Accepted: date}

\maketitle

\begin{abstract}
Graph sampling allows mining a small representative subgraph from a big graph. Sampling algorithms deploy different strategies to replicate the properties of a given graph in the sampled graph. In this study, we provide a comprehensive empirical characterization of five graph sampling algorithms on six properties of a graph including degree, clustering coefficient, path length, global clustering coefficient, assortativity, and modularity. We extract samples from fifteen graphs grouped into five categories including collaboration, social, citation, technological, and synthetic graphs. We provide both qualitative and quantitative results. We find that there is no single method that extracts true samples from a given graph with respect to the properties tested in this work. Our results show that the sampling algorithm that aggressively explores the neighborhood of a sampled node performs better than the others.
\keywords{Graph Sampling \and Graph Properties \and Empirical Characterization}
\end{abstract}

\section{Introduction}
In the last few years, an explosive growth of online networks has attracted millions of users from around the globe. This huge user base provides many opportunities to analyze user behavior \citep{char_user}, social interaction \citep{user_inter}, and information propagation patterns \citep{what_twitter}, to name a few. However, given a large real-world graph with millions of vertices and edges, it is very difficult to apply typical graph processing approaches to analyze the graph directly. As a result, various graph sampling techniques have been proposed for the analysis or mining of large complex networks. Graph sampling is a technique to extract a small subgraph from the original large graph such that the properties of the original graph are preserved in the sample. If true, analyzing the sampled graph should have approximately the same effect as analyzing the original graph.

Many properties have been defined to characterize a graph
and these properties are very important to understand the graph \citep{Survey1}. To our understanding, an ideal representative sample of a large graph should preserve all the properties of a graph as accurately as possible. However, previous studies \citep{walking,statistical} show that some sampling methods are biased towards high degree nodes. It means that such sampling methods tend to sample high degree nodes more often than low degree nodes. In other words, such sampling approaches cannot preserve degree-related properties, e.g., node degree distribution, and at the same time compromise over other properties too. Moreover, some properties, e.g., assortativity and modularity have not been discussed by the research community in association with graph sampling. 


In this paper, we aim at providing a comprehensive empirical characterization of several graph sampling algorithms on a bigger set of properties of a graph. We extract samples from fifteen graphs with five state-of-the-art sampling methods at different sampling fractions.  We characterize the sampling methods on six properties of a graph including three local properties, i.e., degree, clustering coefficient, and path length, and three global properties, i.e., global clustering coefficient, assortativity, and modularity. We provide both qualitative and quantitative results for a better understanding. To the best of our knowledge, we are the first to present such a comprehensive study on graph sampling algorithms. We believe that this study will provide new insights into the graph sampling research and it would be helpful in designing better sampling methods in the future. 

The rest of the paper is organized as follows. In section 2, we provide some definitions and discuss six properties of a graph. In section 3, we overview five state-of-the-art sampling algorithms. In section 4, we discuss the evaluation criteria and the datasets used in this study. In section 5, we present the experimental evaluation and results of our study. In section 6, we discuss related work and conclude the paper in section 7.

\section{Background}
\subsection{Definitions}
Real-world networks can be represented as either directed or undirected graphs. In this work, we study undirected graphs only. An undirected graph is represented as G = (V,E), where V = $\{v_1,v_2,v_3, ... , v_n\}$ is the set of vertices (or nodes) and E = $ \{e_1, e_2, e_3, ... , e_m\}$ is the set of edges (or links). The total number of nodes and edges are represented as $|V| = n$ and $|E| = m$ respectively. A sampling algorithm extracts a sample graph $G_s = (V_s, E_s)$ from G such that $V_s \subset V$ and $E_s \subset E$.  The resulting sample graph $G_s$ has $n_s$ number of vertices and $m_s$ number of edges in it. Given a sampling fraction or sampling budget $\phi$ such that $|V_s|/|V|$ = $\phi$, the aim of sampling is to obtain a sample with a small value of $\phi$ while preserving the properties of $G$ in $G_s$.

\subsection{Graph Properties}
In this paper, we consider three local and three global properties of a graph. A local property assigns a value to every node in a graph. We consider degree, clustering coefficient, and path length as local properties of a graph. A global property does not associate a value to every node in a graph rather it has a value for the whole graph. We consider global clustering coefficient, assortativity, and modularity as global properties of a graph. \\
\textbf{Degree:} The degree of a node is defined as the number of edges connected to the node in the graph while the average degree is simply the average number of edges per node in the graph. We represent the degree of node $v$ as $d_v$ and calculate the average degree $d_{avg}(G)$ of graph $G$ as:
\begin{equation}
	d_{avg}(G) = \frac{2*m}{n}
\end{equation}
The degree distribution $P(d)$ of a graph is defined as the fraction of nodes in the network with degree $d$. Thus if there are $n$ nodes in total in a network and $n_d$ of them have degree $d$, we have 
\begin{equation}
	P(d) = \frac{n_d}{n}
\end{equation}
We find both the average degree and degree distribution of graphs in this work.
\textbf{Clustering Coefficient:} The clustering coefficient measures the average probability that two neighbors of a vertex are themselves neighbors. The local clustering coefficient $c_v$ of node $v$ of degree $d_v$ is the proportion of the number of edges $e_v$ between the neighbors of $v$ relative to the total number of possible edges between the neighbors, given by
\begin{equation}
	c(v) = \frac{2*e_v}{d_v(d_v-1)}
\end{equation}
The average clustering coefficient $c_{avg}$ of a graph is calculated as:
\begin{equation}
	c_{avg} = \frac{1}{n}\sum_{i=1}^{n} c_i
\end{equation}
We also find the distribution of local clustering coefficients to show the fraction of nodes having a particular value of clustering coefficient. \\
\textbf{Path length:} In a graph, path length is defined as the total number of edges traversed while going from a source node to a destination node. The average path length is calculated as the number of edges along the shortest paths for all possible pairs of network nodes. Let $d(v_{i},v_{j})$ where $v_{i},v_{j}\in V$ denote the shortest distance between nodes $v_i$ and $v_j$. Then, the average path length $l_{avg}$ is:
\begin{equation}
l_{avg} = \frac{1}{n(n-1)}\sum_{i\neq j} d(v_i, v_j)
\end{equation}
The frequency distribution of these path lengths define the path length distrubtion. \\
\textbf{Global Clustering Coeffient:} The global clustering coefficient is based on triplets of nodes. A triplet has three nodes that are connected by either two (open triplet) or three (closed triplet) undirected edges. The global clustering coefficient $c_g(G)$ of a graph $G$ is defined as the ratio of the number of closed triplets to the total number of triplets (both open and closed).
\begin{equation}
	c_g(G) = \frac{number\; of\; closed\; triplets}{total\; number\; of\; triplets}
\end{equation}
\\
\textbf{Assortativity:} It measures the tendency of high-degree nodes connect to other high-degree nodes and vice-versa in a network using the Pearson correlation coefficient \citep{Newman}. Positive values of this coefficient indicate a correlation between nodes of similar degrees while negative values indicate relationships between nodes of different degrees. The assortativity coefficient is the Pearson correlation coefficient of degree between pairs of linked nodes. It is given by $r$ as: 
\begin{equation}
	r = \frac{\sum_{jk}jk(e_{jk}-q_jq_k)}{\sigma_q^2}
\end{equation}
where $e_{jk}$ is the joint probability distribution of the remaining degree of the node pair and $q_k$ is the distribution of the remaining degree. When $r = 1$, the network is said to be assortative, when $r = 0$ the network is non-assortative, while at $r = -1$ the network is completely disassortative.\\
\textbf{Modularity:} Modularity is one of the measures of the structure of networks or graphs. It measures the strength of division of a network into modules (clusters or communities). Networks with high modularity have dense connections between the nodes within modules but sparse connections between nodes in different modules. Modularity is the fraction of the edges that fall within the given modules minus the expected fraction if the edges were distributed at random \citep{Newman_Mod}. Given the partition of a network into a set of communities $C_i$, the degree of modularity $Q$ associated to this partition can be measured as follows:

\begin{equation}
	Q = \frac{1}{2m}\sum_{ij}[A_{ij} - \frac{k_ik_j}{2m}]\delta(C_i, C_j)
\end{equation}
where $m$ is the total number of edges in the network, $k_i$ is the degree of node $i$, $A_{ij}$ are the elements of adjacency matrix, $C_i$ is the community $i$th node belongs to and $\delta(x)=1$ if $x>=1$ and $\delta(x)=0$ otherwise.
\section{Graph Sampling}

Graph sampling is a technique to extract a small subgraph $G_s$ from a big graph $G$ such that the subgraph truly represents the original graph. In some scenarios, the whole graph is known in advance and sampling is used to obtain a smaller graph. In other scenarios, the whole graph is unknown and sampling is performed to explore the graph. Although sampled graphs are smaller in size, they are similar to original graphs in some way. In the last decade or so, numerous graph sampling methods have been proposed. These methods can be categorized into Node Sampling, Edge Sampling and Traversal-based Sampling.  Of these categories, Traversal-based Sampling has a long history and has some advantages over the other two \citep{Survey1}. Regardless of the sampling method, we are particularly interested in what graph properties are preserved given a sampling method. If some properties are preserved, we can construct efficient estimators for them.

In this paper, we explore how existing sampling methods
perform in preserving different important properties of original graphs. We explore the following five sampling algorithms for this purpose.\\
\textbf{Frontier Sampling (FS): } In FS \citep{FS}, we deploy m-dimensional dependent random walks to sample a graph. It requires a special estimator function to estimate the metric to remove the bias
introduced by Random Walks. FS works in three steps. First, it randomly chooses a set of nodes, $S$, as seeds. Second, it selects a seed $v$ from $S$ with the probability $P(v)$ given as:
\begin{equation}
	P(v) = \frac{d_v}{\sum_{u \in S} d_u}
\end{equation}
Third, an edge $e(v,w)$ is selected uniformly from the edges incident to $v$. It then adds the edge $e(v,w)$ to the set of sampled edges and replaces $v$ with $w$ in $S$. Similar to other methods \citep{walking}, FS ignores isolated nodes in a graph. It should be noted that in order to study any metric under FS method, we must construct a particular estimator, rather than just studying the sampled nodes and edges directly.\\
\textbf{Expansion Sampling (XS): } The XS strategy \citep{XS} is based on the concept of expansion from work on expander graphs. It was particularly designed to sample communities in networks. It extracts a stratified sample of community structure, i.e., it tries to sample nodes from all the communities in a graph. The algorithm seeks to greedily construct the sample by maximizing the expansion factor $X(S)$ of sample $S$ given by:
\begin{equation}
	X(S)  = \frac{|N(S)|}{|S|}
\end{equation}
where $N(S)$ is the neighborhood of $S$. In our work, we implement Snowball Expansion Sampling as it performs slightly better than Markov
Chain Monte Carlo Expansion Sampling \citep{XS}. \\
\textbf{Rank Degree (RD): } RD \citep{RD} is a graph
exploration method based on the ranking of nodes according
to their degree values. The algorithm takes three parameters $(s,\rho, x)$ as input where $s$ is the number of initial seeds, $0 < \rho \leq 1$  defines the top-k nodes of each ranking list and $x$ is the sample size. The algorithm starts with a set of seed nodes and picks a node at random from this set. It then ranks the neighbors of the selected seed node according to their degree values and selects top-k nodes from the ranked list. The selected top-k nodes along with the edges with the seed node are added to the sample graph. Moreover, these top-k nodes form the set of seeds in the next iteration. \\
\textbf{List Sampling (LS): } The work \citep{LS} claims that the previous methods do not explore the neighborhood of sampled nodes fairly and hence yield sub-optimal samples. It then introduces a new approach in which we keep a list of candidate nodes that is populated with all the neighbors of nodes that have been sampled so far. By doing this, we balance the depth and breadth of graph exploration and produce better samples. The paper proposes three algorithms based on this idea that differ in how to select nodes from the candidate list. We implement LS2 algorithm in this work as it performs better than the other two variations \citep{LS}. \\
\textbf{Hybrid Jump (HJ):} HJ \citep{HJ} introduces a hybrid jump strategy into Metropolis–Hasting Random Walk during the sampling process. It uses a breadth-first search to obtain a set of unique nodes from a list of jump nodes. By applying Uniform Sampling, it gets the average degree of the original network and determines the optimal value of the jump parameter.\\

\section{Evaluation Criteria}
We evaluate the above mentioned five sampling methods both qualitatively and quantitatively against one another. We evaluate them on both real-world and synthetic graphs and see how well they preserve the above mentioned six properties of a graph. 

\subsection{Datasets}
We perform experiments on both real-world and synthetic datasets. We use 12 real-world and 3 synthetic datasets. The real datasets are drawn from a wide range of networks including collaboration networks, social networks, citation networks and topological networks. The size of these networks vary from a few thousands to a million of nodes. All these datasets are publicly available \citep{Sdata,Ndata,Kdata}.

Moreover, we also extract samples from three synthetic networks because these networks have strong mathematical foundations and it would be interesting to sample them. We select three generative models for synthesizing networks and run the sampling methods on the generated networks. The selection of these models is based on the fact that they generate graphs that follow many properties of real-world graphs. The parameter values of each generative model are tuned such that the generated network has nearly the same number of nodes and edges as that of the average of the twelve real-world datasets. The three generative models are: 1) Forest Fire (FF) \citep{FF_Gen} 2) Small World (SW) \citep{SW_Gen} and 3) Mixed Model (MM) \citep{MM_Gen}.
We use sampling fractions $\phi$ = \{0.02, 0.04, 0.06, 0.08, 0.1\} to extract samples from these graphs. Table~\ref{Tab_datasets} summarizes the characteristics of these datasets.

\begin{table} 
	\centering
		\captionof{table}{Characteristics of the datasets used in the experiments}
		\scalebox{0.85}{
\begin{tabular}{|l|l|l|p{1.0cm}|p{1.0cm}|p{1.0cm}|p{1.0cm}|p{1.0cm}|p{1.0cm}|p{1.7cm}|}
	\hline
	\textbf{Dataset} & \textbf{Nodes} & \textbf{Edges} & \textbf{Avg. Degree} & \textbf{Avg. Clust. Coeff.} & \textbf{Avg. Path length} & \textbf{Assort-ativity} & \textbf{Global Clust. Coeff.} & \textbf{Modul-arity} & \textbf{Network Type} \\ \hline
	CiteSeer & 227,320 & 814,134 & 7.16 & 0.76 & 7.82 & 0.07 & 0.45 & 0.89 & Collaboration \\[1mm] 
	DBLP & 317,080 & 1,049,866 & 6.62 & 0.73 & 6.79 & 0.26 & 0.31 & 0.81 & Collaboration \\[1mm] 
	Actors & 382,219 & 15,038,084 & 78.68 & 0.78 & 3.56 & 0.22 & 0.16 & 0.68 & Collaboration \\ [1mm]
	Gowalla & 196,591 & 950,327 & 9.66 & 0.31 & 4.62 & -0.03 & 0.02 & 0.69 & Social \\ [1mm]
	Digg & 770,799 & 5,907,132 & 15.32 & 0.16 & 4.49 & -0.09 & 0.05 & 0.53 & Social \\ [1mm]
	Hyves & 1,402,673 & 2,777,419 & 3.96 & 0.11 & 5.67 & -0.02 & 0.01 & 0.76 & Social \\ [1mm]
	Cora & 23,166 & 89,157 & 7.69 & 0.31 & 5.74 & -0.05 & 0.12 & 0.78 & Citation \\ [1mm]
	HepTh & 27,769 & 352,285 & 25.37 & 0.32 & 4.27 & -0.03 & 0.11 & 0.64 & Citation \\ [1mm]
	HepPh & 34,546 & 420,877 & 24.36 & 0.29 & 4.41 & -0.01 & 0.14 & 0.72 & Citation \\ [1mm]
	Topology & 34,761 & 107,720 & 6.19 & 0.42 & 3.78 & -0.21 & 0.05 & 0.61 & Technological  \\ [1mm]
	Gnutella & 62,586 & 147,892 & 4.72 & 0.01 & 5.96 & -0.09 & 0.01 & 0.49 & Technological  \\ [1mm]
	Caida & 190,914 & 607,610 & 6.36 & 0.21 & 6.98 & 0.02 & 0.06 & 0.83 & Technological  \\ [1mm]
	FF & 300,000 & 2,446,862 & 16.31 & 0.16 & 3.34 & -0.06 & 0.01 & 0.27 & Synthetic \\ [1mm]
	SW & 300,000 & 2,400,000 & 16.00 & 0.37 & 5.89 & 0.00 & 0.36 & 0.78 & Synthetic \\ [1mm]
	MM & 300,000 & 2,322,676 & 15.48 & 0.39 & 3.44 & -0.17 & 0.07 & 0.15 & Synthetic \\ \hline
\end{tabular}}
	\label{Tab_datasets}
\end{table} 

\subsection{Evaluation Metrics}
We evaluate the quality of samples by comparing the above mentioned properties of sample graphs with that of the original graphs. We also perform quantitative tests  such as Jensen-Shannon Distance (JSD) and Root Mean Square Error (RMSE) for quantitative evaluation of sampling algorithms. All the results presented in this paper are averaged over ten readings. We use the following metrics for evaluation.\\
\textbf{Point Statistics:} A point statistic shows the value of a property at a single point.  We vary the sampling fraction $\phi$ from 0.02 to 0.1  and plot the scaling ratio of a property $\Theta$ as the ratio of the value of that property in the sampled graph $\Theta_S$ to the value of that property in the actual graph $\Theta_A$:

\begin{equation}
	\,Scaling\, Ratio=\frac{\Theta_S}{\Theta_A}
\end{equation}
For example, we measure the average degree of the sample and original graphs and find the scaling ratio of degree by diving the average degree of the sample graph $G_s$ at a sampling fraction $\phi$ with the average degree of the original graph $G$. \\
\textbf{Distributions:} A distribution is a multivalued statistic and shows the distribution of a property in a graph. For example, the degree distribution shows the fraction of nodes that have degree greater than or less than a particular value. We find and plot the Empirical Cumulative Distribution Function (ECDF) of degree, clustering coefficient and path lengths of sample graphs at $\phi$ = 0.02.\\
\textbf{Root Mean Square Error:} Given the original graph $G$ and sampled graph $G_s$, we want to measure how far is $G_s$ from G. For scalar quantities such as the average degree, we use the common measure for the quality of estimation by Root Mean Square Error (RMSE), given as
\begin{equation}
	\, RMSE = \sqrt{\frac{1}{n}\sum_{1}^{n} (\Theta_S - \Theta_A)^2}
\end{equation}
where $\Theta_S$ and $\Theta_A$ are sampled and original values respectively.\\
\textbf{Jensen-Shannon Distance:} For distributions of the properties, we measure the Jensen-Shannon Distance. In probability theory, the Jensen-Shannon Divergence measures the similarity between two probability distributions, calculated as
\begin{equation}
	\, D_{JS}(P||Q) = \frac{1}{2}D_{KL}(P||M) + \frac{1}{2}D_{KL}(Q||M)
\end{equation}
where $D_{JS}$ and $D_{KL}$ are Jensen-Shannon and Kullback-Leibler Divergences respectively while P and Q are two Probability Distribution Functions (PDFs) and M = $\frac{1}{2}$(P+Q). Its square root is a true metric often referred to as Jensen-Shannon Distance (JSD).

\section{Experimental Evaluation and Analysis}

In this section, we evaluate FS, XS, RD, LS and HJ sampling algorithms in terms of six graph properties defined above. We analyze the performance of each algorithm and consider the properties of the original graphs as ground truth values. 

\subsection{Degree}
\begin{figure}[H]
	\centering
	\includegraphics[width = 120mm, height=150mm]{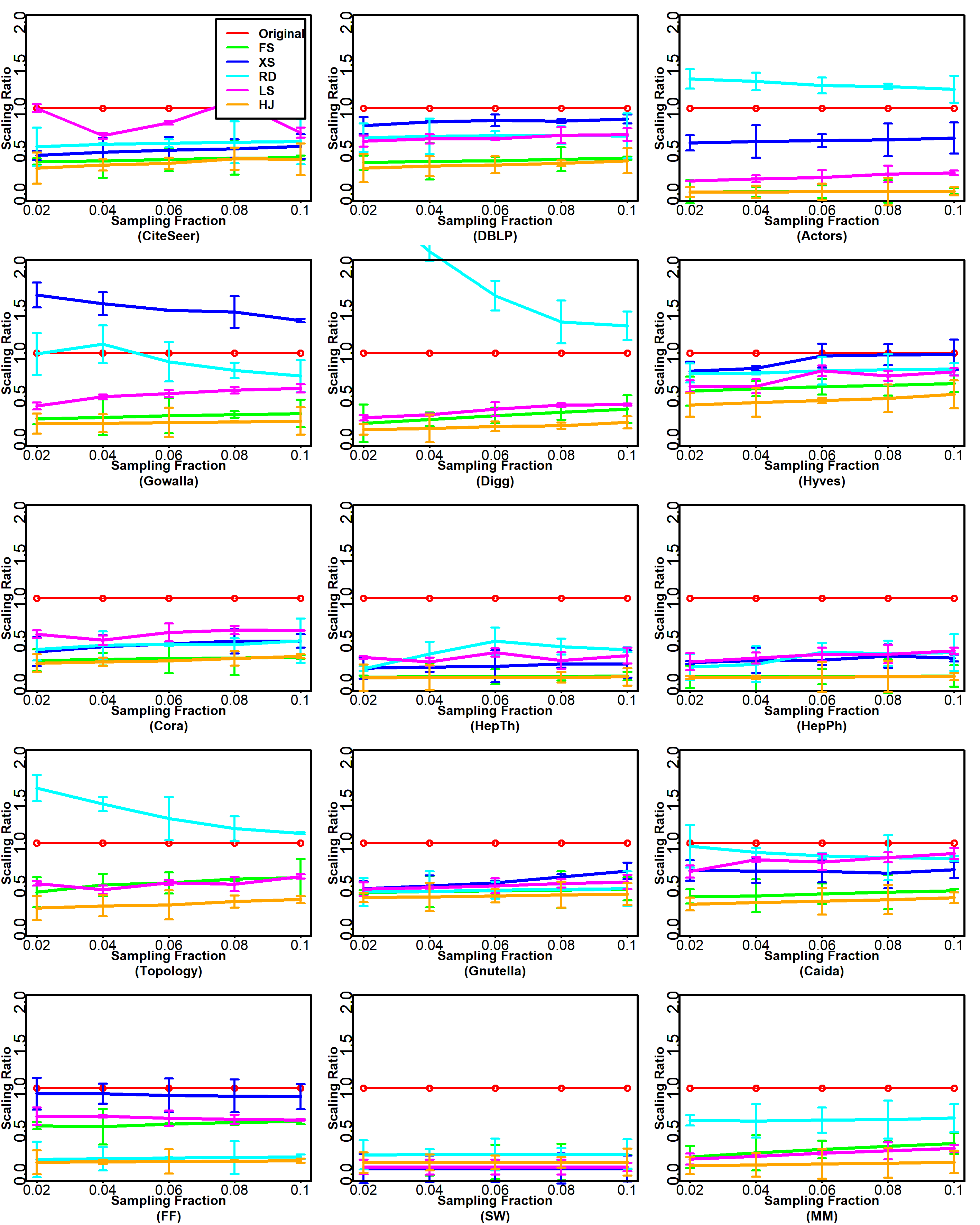}
	\caption{Point statistics of average degree of all the networks with 95\% confidence intervals.}
	\label{fig_Deg_Stats}	
	\vspace{0.75cm}
	\captionof{table}{RMSE values and standard deviations of point statistics of average degree. \textbf{Boldface} values are the best results.}
	\scalebox{0.95}{
			\begin{tabular}{|l|l|l|l|l|l|}
			\hline
			Datasets & \textbf{FS} & \textbf{XS} & \textbf{RD} & \textbf{LS} & \textbf{HJ} \\ \hline
			CiteSeer & 4.27$\pm$0.21 & 3.52$\pm$0.16 & 2.95$\pm$0.12 & \textbf{1.23}$\pm$0.10 & 4.55$\pm$0.16 \\ 
			DBLP & 4.02$\pm$0.14 & \textbf{1.02}$\pm$0.03 & 2.15$\pm$0.20 & 2.27$\pm$0.11 & 4.34$\pm$0.10 \\ 
			Actors & 76.30$\pm$0.18 & 29.60$\pm$0.12 & \textbf{21.82}$\pm$0.15 & 62.81$\pm$0.08 & 76.43$\pm$0.07 \\ 
			Gowalla & 7.05$\pm$0.26 & 5.04$\pm$0.12 & \textbf{1.30}$\pm$0.23 & 4.69$\pm$0.02 & 7.78$\pm$0.09 \\ 
			Digg & 11.16$\pm$0.24 & 31.63$\pm$0.04 & 13.39$\pm$0.08 & \textbf{10.17}$\pm$0.16 & 13.06$\pm$0.19 \\ 
			Hyves & 1.56$\pm$0.10 & \textbf{0.35}$\pm$0.03 & 0.82$\pm$0.11 & 1.15$\pm$0.04 & 2.16$\pm$0.07 \\ 
			Cora & 5.40$\pm$0.25 & 4.17$\pm$0.21 & 4.18$\pm$0.15 & \textbf{3.15}$\pm$0.11 & 5.53$\pm$0.06 \\ 
			HepTh & 23.13$\pm$0.13 & 19.97$\pm$0.11 & \textbf{15.90}$\pm$0.19 & 17.51$\pm$0.19 & 23.34$\pm$0.08 \\ 
			HepPh & 22.17$\pm$0.33 & 17.33$\pm$0.07 & \textbf{16.91}$\pm$0.09 & 16.30$\pm$0.03 & 22.32$\pm$0.14 \\ 
			Topology & 2.89$\pm$0.00 & 15.33$\pm$0.11 & \textbf{2.06}$\pm$0.24 & 2.88$\pm$0.00 & 4.38$\pm$0.16 \\ 
			Gnutella & 2.60$\pm$0.18 & \textbf{2.08}$\pm$0.11 & 2.59$\pm$0.02 & 2.32$\pm$0.08 & 2.89$\pm$0.05 \\ 
			Caida & 3.74$\pm$0.13 & 2.06$\pm$0.32 & \textbf{0.80}$\pm$0.02 & 1.30$\pm$0.13 & 4.28$\pm$0.11 \\
			FF & 6.81$\pm$0.21 & \textbf{1.30}$\pm$0.07 & 13.25$\pm$0.00 & 5.65$\pm$0.14 & 13.89$\pm$0.07 \\ 
			SW & 13.77$\pm$0.25 & 14.95$\pm$0.06 & \textbf{12.32}$\pm$0.18 & 14.67$\pm$0.01 & 13.81$\pm$0.02 \\ 
			MM & 11.09$\pm$0.30 & 15.39$\pm$0.16 & \textbf{5.68}$\pm$0.22 & 11.76$\pm$0.18 & 13.63$\pm$0.09 \\ \hline
			Average &13.06$\pm$0.19	&20.25$\pm$0.12	&\textbf{7.74}$\pm$0.13	&10.52$\pm$0.09	&14.16$\pm$0.10
			\\ \hline
			
		\end{tabular}
	 }
	\label{Tab_Degree_RMSE}
\end{figure}

In the first experiment, we vary the sampling fraction from $\phi=0.02$ to $\phi=0.1$  and plot the scaling ratio of the average degree in Figure~\ref{fig_Deg_Stats} for all the graphs. All the values are shown with 95\% confidence intervals. The general trend is that all the algorithms under-sample the graphs with a few exceptions and produce samples with an average degree less than that of the corresponding original graphs. The possible reason is that most of these algorithms, i.e., FS, XS, and HJ do not explicitly prefer high degree nodes and it seems that they sample low degree nodes more often than high degree nodes. RD prefers high degree nodes but the initial seed nodes are defined randomly, allowing the algorithm to start from different areas of the graph, where some nodes may be of low degree. Since RD and LS tend to collect high degree nodes more often, therefore, they perform slightly better than the other methods. Table~\ref{Tab_Degree_RMSE} gives the RMSE values of all the methods for each dataset where a value is averaged over all the sampling fractions. The table also shows the standard deviation values. We see that, on average, RD samples produce minimum error among the tested algorithms whereas LS stands second to it. All the methods perform the worst in the Actors dataset and a possible reason is the very high degree of this dataset. Similarly, the samples of HepTh and HepPh datasets also produce big error values because of the high degree of the original graphs.

\begin{figure}[H]
	\centering
	\includegraphics[width = 120mm, height=150mm]{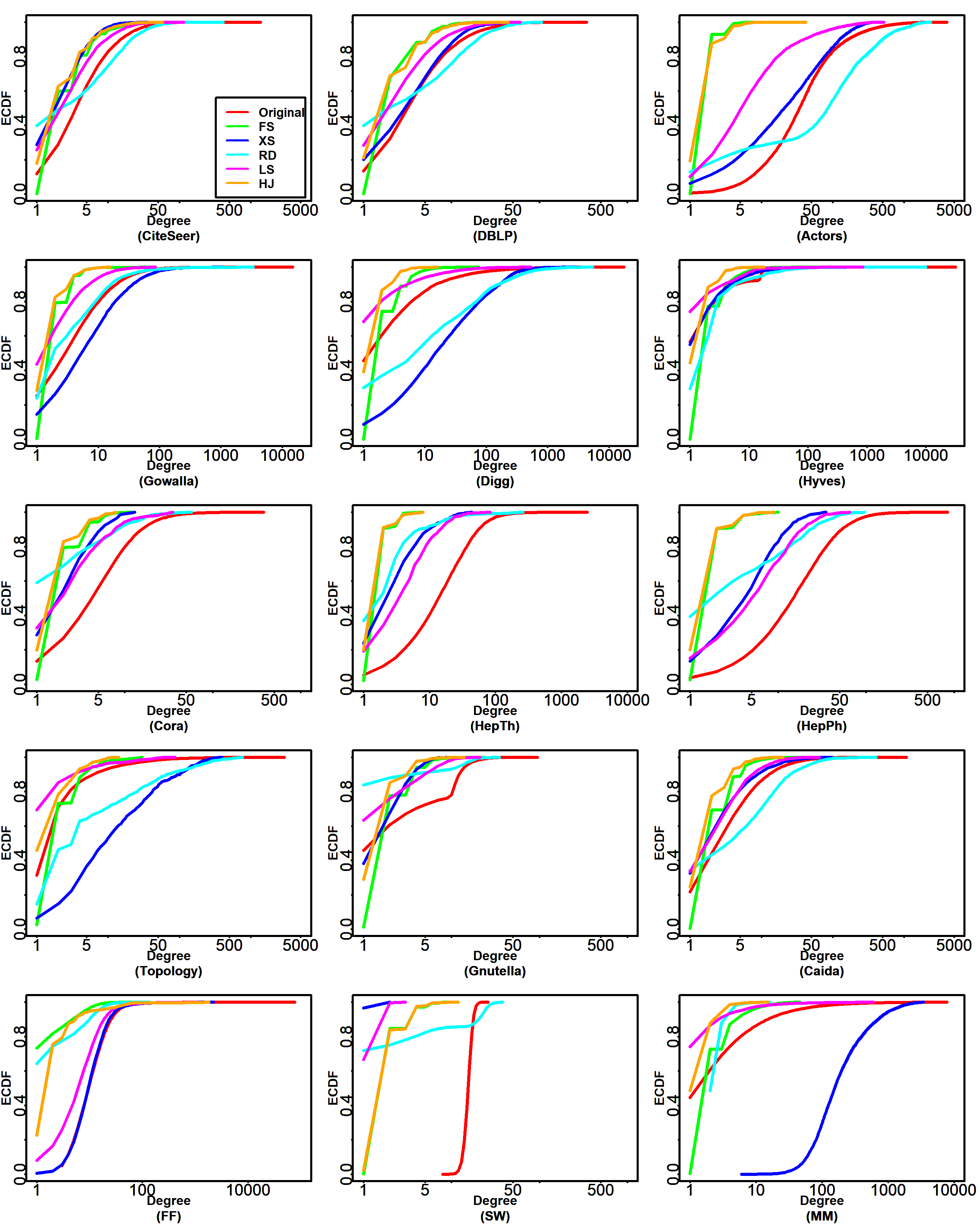}
	\caption{Degree distributions of all the networks at $\phi=0.02$. \textbf{(Best viewed in color.)} }
	\label{fig_Deg_Dist}
	\vspace{0.75cm}
	\captionof{table}{Jensen Shannon distance and standard deviations for degree distributions. \textbf{Boldface} values are the best results.}
	\scalebox{0.95}{
			\begin{tabular}{|l|l|l|l|l|l|}
			\hline
			Datasets & \textbf{FS} & \textbf{XS} & \textbf{RD} & \textbf{LS} & \textbf{HJ} \\ \hline
			CiteSeer & 0.44$\pm$0.05 & 0.22$\pm$0.04 & 0.29$\pm$0.01 & \textbf{0.15}$\pm$0.02 & 0.29$\pm$0.03 \\ 
			DBLP & 0.41$\pm$0.02 & \textbf{0.08}$\pm$0.02 & 0.28$\pm$0.01 & 0.16$\pm$0.09 & 0.31$\pm$0.06 \\ 
			Actors & 0.56$\pm$0.10 & \textbf{0.20}$\pm$0.04 & 0.42$\pm$0.05 & 0.42$\pm$0.02 & 0.56$\pm$0.11 \\ 
			Gowalla & 0.51$\pm$0.11 & 0.15$\pm$0.05 & \textbf{0.13}$\pm$0.00 & 0.18$\pm$0.06 & 0.33$\pm$0.01 \\ 
			Digg & 0.56$\pm$0.05 & 0.40$\pm$0.07 & 0.27$\pm$0.06 & \textbf{0.16}$\pm$0.02 & 0.27$\pm$0.02 \\ 
			Hyves & 0.59$\pm$0.05 & \textbf{0.12}$\pm$0.01 & 0.25$\pm$0.10 & 0.19$\pm$0.06 & 0.23$\pm$0.01 \\ 
			Cora & 0.48$\pm$0.01 & 0.24$\pm$0.09 & 0.34$\pm$0.06 & \textbf{0.23}$\pm$0.04 & 0.38$\pm$0.01 \\ 
			HepTh & 0.51$\pm$0.04 & 0.34$\pm$0.12 & 0.41$\pm$0.05 & \textbf{0.27}$\pm$0.05 & 0.47$\pm$0.12 \\ 
			HepPh & 0.53$\pm$0.01 & 0.30$\pm$0.01 & 0.37$\pm$0.06 & \textbf{0.25}$\pm$0.03 & 0.49$\pm$0.01 \\ 
			Topology & 0.39$\pm$0.08 & 0.45$\pm$0.09 & 0.28$\pm$0.03 & 0.27$\pm$0.01 & \textbf{0.11}$\pm$0.09 \\ 
			Gnutella & 0.56$\pm$0.04 & \textbf{0.18}$\pm$0.10 & 0.30$\pm$0.06 & 0.23$\pm$0.01 & 0.34$\pm$0.05 \\ 
			Caida & 0.45$\pm$0.02 & \textbf{0.10}$\pm$0.07 & 0.23$\pm$0.10 & 0.11$\pm$0.05 & 0.29$\pm$0.10 \\ 
			FF & 0.39$\pm$0.07 & \textbf{0.06}$\pm$0.06 & 0.34$\pm$0.12 & 0.13$\pm$0.04 & 0.43$\pm$0.03 \\ 
			SW & 0.08$\pm$0.01 & 0.62$\pm$0.11 & 0.55$\pm$0.11 & 0.63$\pm$0.05 & \textbf{0.07}$\pm$0.03 \\ 
			MM & 0.53$\pm$0.02 & 0.53$\pm$0.03 & 0.34$\pm$0.00 & \textbf{0.22}$\pm$0.11 & 0.24$\pm$0.11 \\ \hline
			Average & 0.47$\pm$0.05 & 0.27$\pm$0.06 & 0.32$\pm$0.06 & \textbf{0.24}$\pm$0.04 & 0.32$\pm$0.05 \\ \hline
	\end{tabular} }
	\label{Tab_JS_Degree}
	
\end{figure}

We also find the distribution of the degrees in the sampled graphs at sampling fraction $\phi$=0.02 and present the results in Figure~\ref{fig_Deg_Dist} in the form of the Empirical Cumulative Distribution Function (ECDF). In general, in the collaboration, social and technological graphs, all the algorithms extract better samples than in the citation networks. The overall trend is that these algorithms tend to pick low degree nodes more often than high degree nodes. We present the Jensen-Shannon distance in Table~\ref{Tab_JS_Degree}. The results show that the XS and JS samples produce less deviation from the original distributions. On average, LS outperforms other methods in this metric. We observe that these algorithms perform better in sampling the collaboration and social graphs whereas performing worse in the citation networks. Possibly, the high average degree of the citation networks could not be achieved with a small sampling budget. In the case of synthetic networks, the algorithms extract better samples from FF and MM graphs.

\subsection{Clustering Coefficient}
\begin{figure}[H]
	\centering
	\includegraphics[width = 120mm, height=150mm]{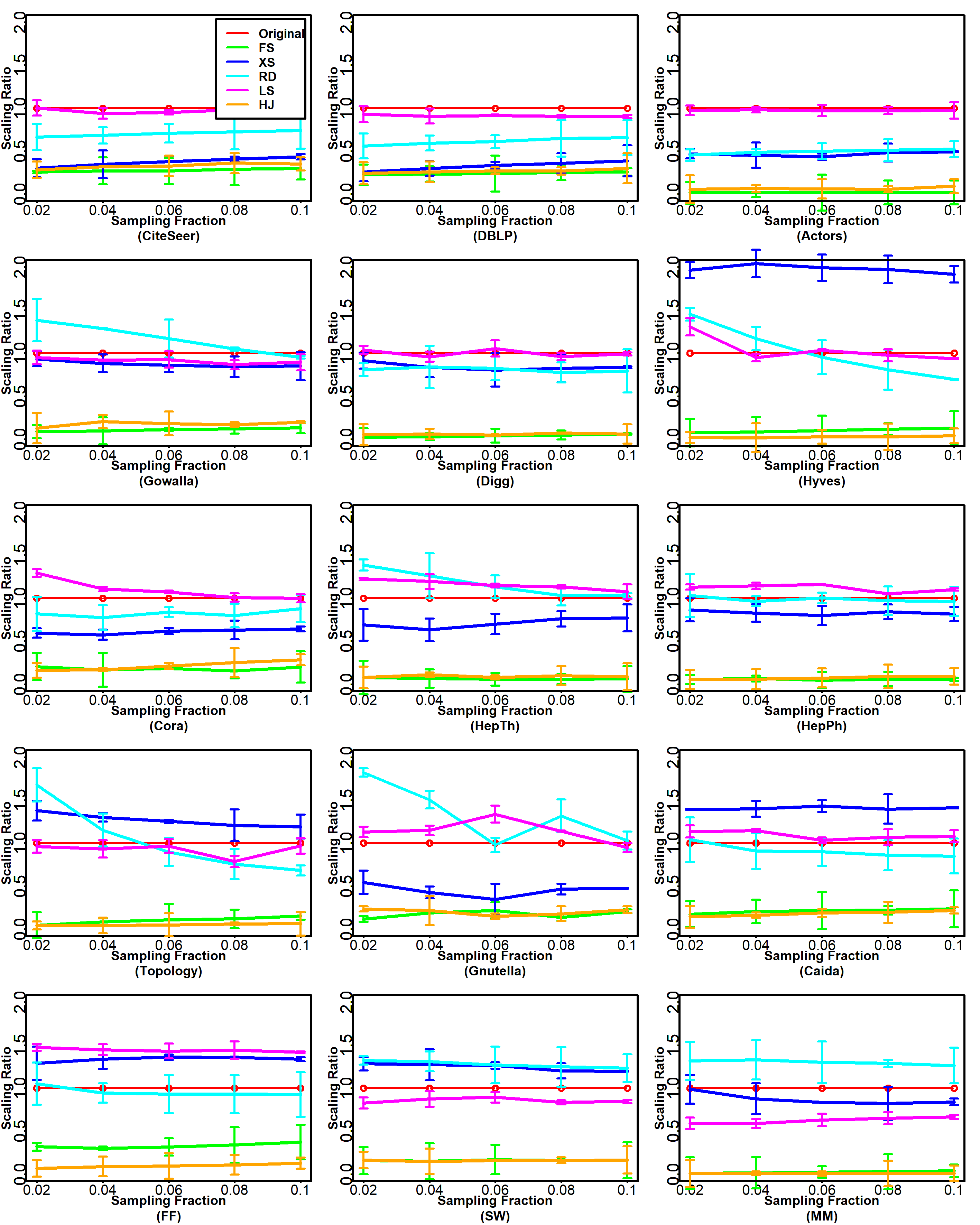}
	\caption{Point statistics of average clustering coefficient of all the networks with 95\% confidence intervals.}
	\label{fig_CC_Stats}
	\vspace{0.75cm}
	\captionof{table}{RMSE values and standard deviations of point statistics of average clustering coefficient. \textbf{Boldface} values are the best results.}	
	\begin{tabular}{|l|l|l|l|l|l|}
		\hline
		Datasets & \textbf{FS} & \textbf{XS} & \textbf{RD} & \textbf{LS} & \textbf{HJ} \\ \hline
		CiteSeer & 0.55$\pm$0.31 & 0.47$\pm$0.22 & 0.22$\pm$0.34 & \textbf{0.02}$\pm$0.03 & 0.51$\pm$0.26 \\ 
		DBLP & 0.55$\pm$0.31 & 0.49$\pm$0.25 & 0.28$\pm$0.19 & \textbf{0.06}$\pm$0.02 & 0.53$\pm$0.06 \\ 
		Actors & 0.77$\pm$0.01 & 0.41$\pm$0.16 & 0.39$\pm$0.20 & \textbf{0.01}$\pm$0.00 & 0.73$\pm$0.09 \\ 
		Gowalla & 0.28$\pm$0.18 & 0.04$\pm$0.21 & 0.05$\pm$0.21 & \textbf{0.02}$\pm$0.08 & 0.26$\pm$0.02 \\ 
		Digg & 0.15$\pm$0.04 & 0.02$\pm$0.24 & 0.03$\pm$0.18 & \textbf{0.00}$\pm$0.01 & 0.15$\pm$0.30 \\ 
		Hyves & 0.09$\pm$0.29 & 0.10$\pm$0.11 & 0.02$\pm$0.24 & \textbf{0.00}$\pm$0.07 & 0.10$\pm$0.14 \\ 
		Cora & 0.25$\pm$0.24 & 0.11$\pm$0.27 & 0.05$\pm$0.05 & \textbf{0.02}$\pm$0.04 & 0.24$\pm$0.17 \\ 
		HepTh & 0.30$\pm$0.30 & 0.09$\pm$0.11 & \textbf{0.05}$\pm$0.09 & 0.05$\pm$0.10 & 0.29$\pm$0.29 \\ 
		HepPh & 0.27$\pm$0.17 & 0.05$\pm$0.18 & \textbf{0.00}$\pm$0.21 & 0.03$\pm$0.03 & 0.27$\pm$0.10 \\ 
		Topology & 0.37$\pm$0.32 & 0.11$\pm$0.27 & 0.12$\pm$0.16 & \textbf{0.03}$\pm$0.11 & 0.39$\pm$0.27 \\ 
		Gnutella & 0.00$\pm$0.11 & \textbf{0.00}$\pm$0.01 & 0.00$\pm$0.13 & 0.00$\pm$0.10 & 0.00$\pm$0.30 \\ 
		Caida & 0.15$\pm$0.41 & 0.08$\pm$0.14 & 0.02$\pm$0.26 & \textbf{0.01}$\pm$0.10 & 0.16$\pm$0.34 \\ 
		FF & 0.10$\pm$0.06 & 0.05$\pm$0.21 & \textbf{0.01}$\pm$0.11 & 0.07$\pm$0.06 & 0.14$\pm$0.00 \\ 
		SW & 0.31$\pm$0.32 & 0.09$\pm$0.12 & 0.10$\pm$0.28 & \textbf{0.05}$\pm$0.09 & 0.31$\pm$0.33 \\ 
		MM & 0.38$\pm$0.22 & \textbf{0.05}$\pm$0.01 & 0.11$\pm$0.06 & 0.14$\pm$0.11 & 0.39$\pm$0.22 \\ \hline
		Average & 0.30$\pm$0.22 & 0.14$\pm$0.17 & 0.10$\pm$0.18 & \textbf{0.04}$\pm$0.06 & 0.30$\pm$0.19 \\ \hline
	\end{tabular}
	\label{Tab_CC_RMSE}	
\end{figure}

We vary the sampling fraction from $\phi=0.02$ to $\phi=0.1$  and plot the scaling ratios of the average clustering coefficient with 95\% confidence intervals in Figure~\ref{fig_CC_Stats}. LS outperforms in the collaboration and social graphs whereas RD stands second in these graphs. Other algorithms underestimate the values of the average clustering coefficient in these networks. It seems that the preference to sample high degree nodes in LS and RD plays its role and enables these methods to preserve even high values of clustering coefficient in the networks. In the citation networks, XS, RD, and LS give comparable results. In the technological and synthetic graphs, both RD and LS produce good samples. Table~\ref{Tab_CC_RMSE} summarizes the RMSE values of all the algorithms in all the datasets. We see that LS gives minimum error in most of the real graphs and outperforms other methods on average in this metric.  RD also gives minimum error in a few datasets and stands second to LS on average.

\begin{figure}[H]
	\centering
	\includegraphics[width = 120mm, height=150mm]{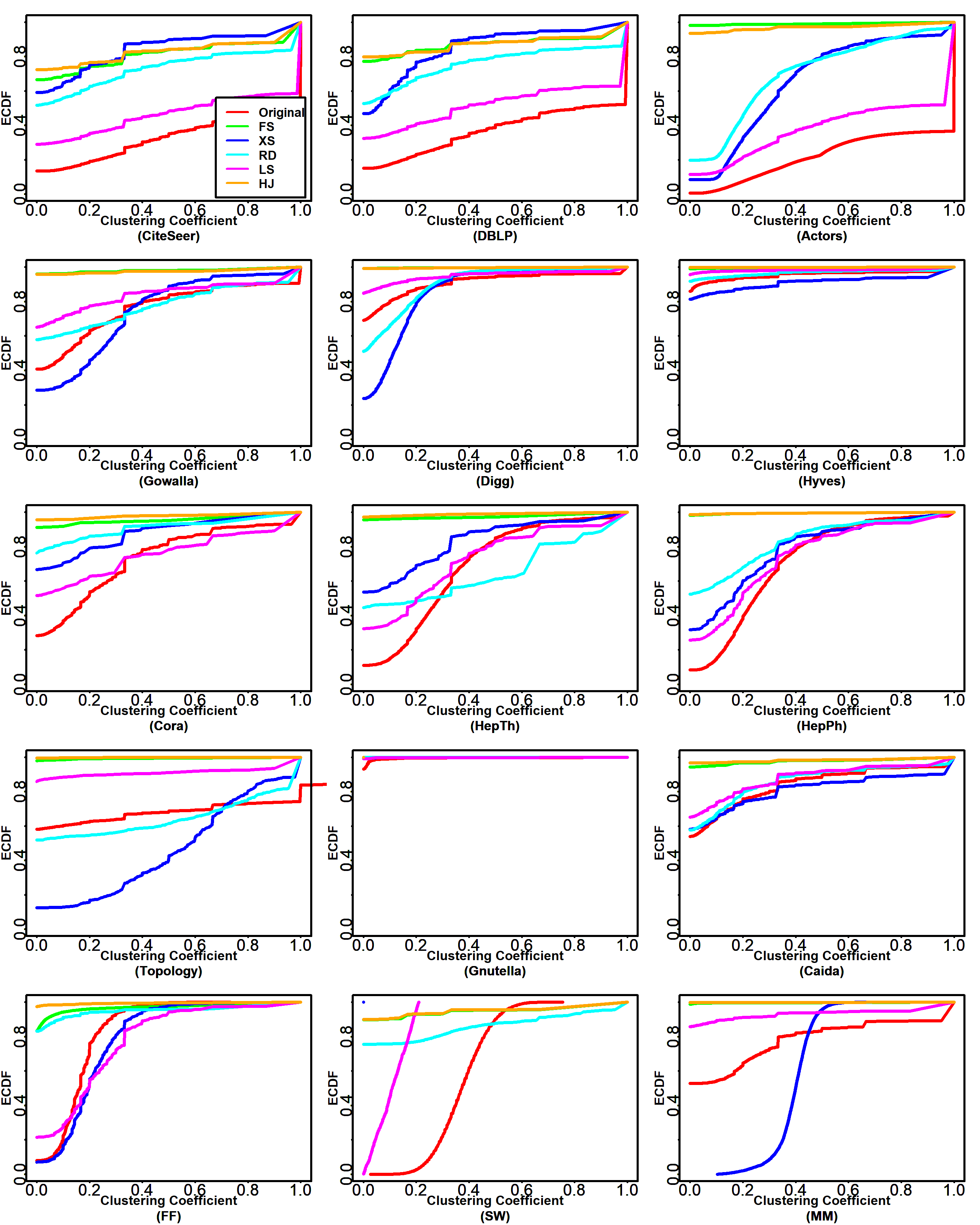}
	\caption{Clustering coefficient distributions of all the networks at $\phi=0.02$. \textbf{(Best viewed in color.)} }
	\label{fig_CC_Dist}
	\vspace{0.75cm}
	\captionof{table}{Jensen Shannon distance and standard deviations for clustering coefficient distributions. \textbf{Boldface} values are the best results.}
	\scalebox{0.95}{
		\begin{tabular}{|l|l|l|l|l|l|}
			\hline
			Datasets & \textbf{FS} & \textbf{XS} & \textbf{RD} & \textbf{LS} & \textbf{HJ} \\ \hline
			CiteSeer & 0.43$\pm$0.06 & 0.45$\pm$0.18 & 0.34$\pm$0.07 & \textbf{0.14}$\pm$0.10 & 0.45$\pm$0.04 \\ 
			DBLP & 0.48$\pm$0.19 & 0.45$\pm$0.10 & 0.34$\pm$0.15 & \textbf{0.15}$\pm$0.01 & 0.49$\pm$0.17 \\ 
			Actors & 0.79$\pm$0.16 & 0.47$\pm$0.16 & 0.52$\pm$0.10 & \textbf{0.20}$\pm$0.17 & 0.75$\pm$0.09 \\ 
			Gowalla & 0.45$\pm$0.10 & 0.19$\pm$0.10 & \textbf{0.18}$\pm$0.03 & 0.19$\pm$0.03 & 0.45$\pm$0.12 \\ 
			Digg & 0.32$\pm$0.10 & 0.36$\pm$0.07 & 0.19$\pm$0.03 & \textbf{0.13}$\pm$0.09 & 0.32$\pm$0.12 \\ 
			Hyves & 0.19$\pm$0.18 & 0.08$\pm$0.08 & \textbf{0.08}$\pm$0.09 & 0.12$\pm$0.08 & 0.21$\pm$0.17 \\ 
			Cora & 0.50$\pm$0.06 & 0.29$\pm$0.17 & 0.38$\pm$0.15 & \textbf{0.21}$\pm$0.01 & 0.53$\pm$0.08 \\ 
			HepTh & 0.67$\pm$0.00 & 0.36$\pm$0.18 & 0.45$\pm$0.18 & \textbf{0.23}$\pm$0.17 & 0.67$\pm$0.08 \\ 
			HepPh & 0.70$\pm$0.19 & 0.26$\pm$0.12 & 0.37$\pm$0.01 & \textbf{0.21}$\pm$0.05 & 0.71$\pm$0.18 \\ 
			Topology & 0.37$\pm$0.12 & 0.47$\pm$0.15 & \textbf{0.18}$\pm$0.08 & 0.24$\pm$0.00 & 0.40$\pm$0.06 \\ 
			Gnutella & 0.15$\pm$0.13 & 0.14$\pm$0.16 & 0.14$\pm$0.09 & \textbf{0.12}$\pm$0.03 & 0.15$\pm$0.05 \\ 
			Caida & 0.34$\pm$0.04 & 0.09$\pm$0.13 & \textbf{0.05}$\pm$0.17 & 0.08$\pm$0.16 & 0.38$\pm$0.16 \\ 
			FF & 0.63$\pm$0.01 & \textbf{0.17}$\pm$0.05 & 0.63$\pm$0.09 & 0.29$\pm$0.06 & 0.71$\pm$0.03 \\ 
			SW & 0.77$\pm$0.07 & 0.80$\pm$0.17 & \textbf{0.68}$\pm$0.00 & 0.74$\pm$0.08 & 0.77$\pm$0.11 \\ 
			MM & 0.42$\pm$0.12 & 0.64$\pm$0.12 & 0.43$\pm$0.03 & \textbf{0.29}$\pm$0.16 & 0.43$\pm$0.12 \\ \hline
			Average & 0.48$\pm$0.10 & 0.35$\pm$0.13 & 0.33$\pm$0.09 & \textbf{0.22}$\pm$0.08 & 0.49$\pm$0.11 \\ \hline
	\end{tabular} }
	\label{Tab_JS_CC}
	
\end{figure}

Figure~\ref{fig_CC_Dist} shows the distribution of clustering coefficients in all the datasets sampled at $\phi=0.02$ by the sampling algorithms. We see that, generally, the samples extracted by LS in the collaboration, social, and citation networks better match with their original counterparts. In the technological networks, LS, XS, and RD extract good samples, however, all the algorithms fail in sampling the small-world network. Besides LS, other methods also extract reasonable samples in few datasets, e.g., XS in the citation networks and RD in the social networks. The JS distance is summarized in Table~\ref{Tab_JS_CC}. On average, LS samples give minimum error and outperform in nine datasets whereas RD stands second to it. It seems that the sampling algorithms find it hard to sample the synthetic networks while they perform relatively better in the other networks.

\subsection{Path length}

\begin{figure}[H]
	\centering
	\includegraphics[width = 120mm, height=150mm]{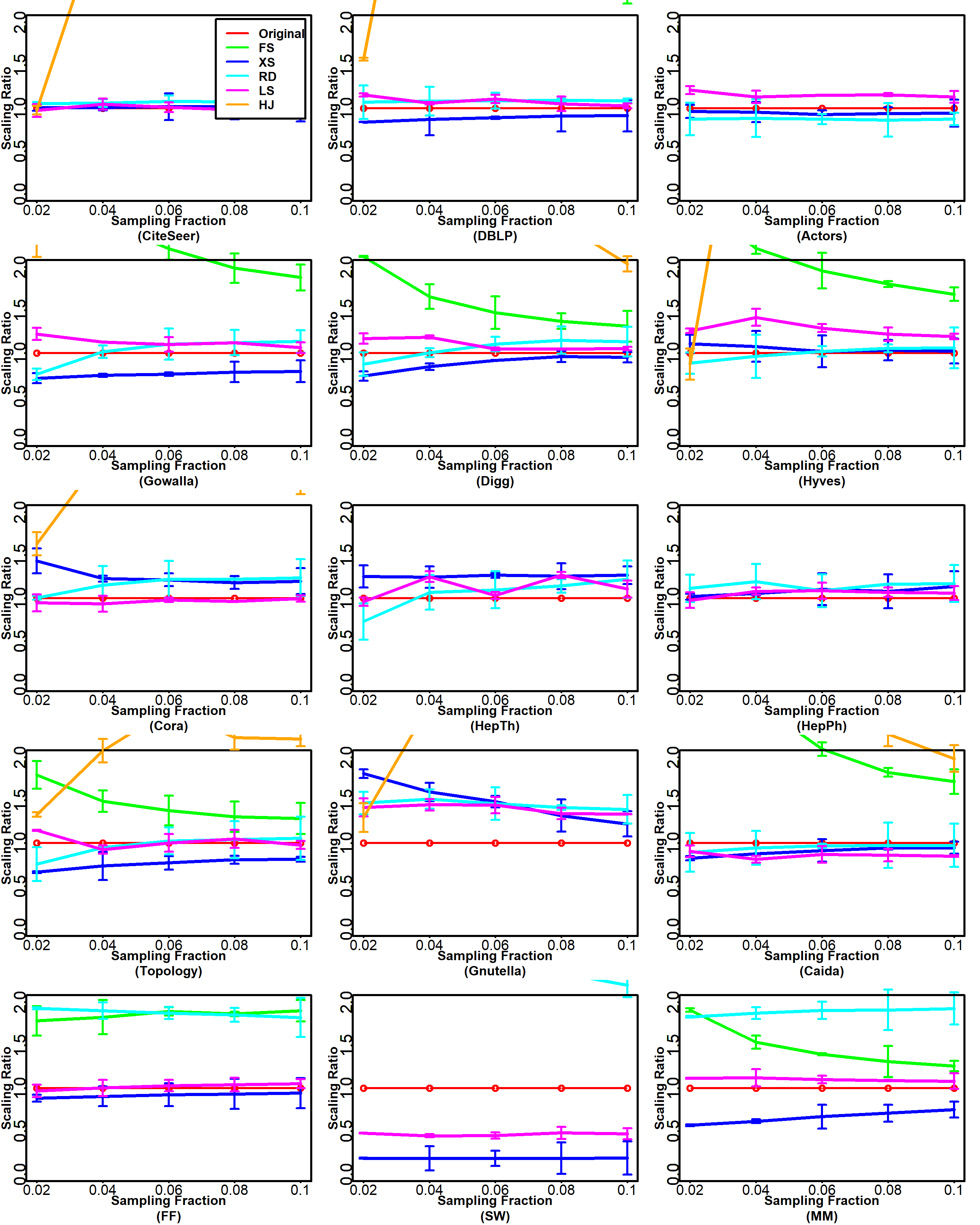}
	\caption{Point statistics of average path length of all the networks with 95\% confidence intervals.}
	\label{fig_PL_Stats}
	\vspace{0.75cm}
	\captionof{table}{RMSE values and standard deviations of point statistics of average path length. \textbf{Boldface} values are the best results.}	
	\begin{tabular}{|l|l|l|l|l|l|}
		\hline
		Datasets & \textbf{FS} & \textbf{XS} & \textbf{RD} & \textbf{LS} & \textbf{HJ} \\ \hline
		CiteSeer & 23.11$\pm$0.19 & \textbf{0.10}$\pm$0.06 & 0.53$\pm$0.21 & 0.18$\pm$0.21 & 12.12$\pm$0.21 \\ 
		DBLP & 18.73$\pm$0.02 & 0.78$\pm$0.04 & 0.56$\pm$0.03 & \textbf{0.53}$\pm$0.03 & 16.59$\pm$0.15 \\ 
		Actors & 18.01$\pm$0.15 & \textbf{0.19}$\pm$0.14 & 0.45$\pm$0.08 & 0.55$\pm$0.04 & 25.82$\pm$0.22 \\ 
		Gowalla & 6.59$\pm$0.11 & 1.12$\pm$0.07 & \textbf{0.57}$\pm$0.05 & 0.59$\pm$0.06 & 17.27$\pm$0.16 \\
		Digg & 2.63$\pm$0.21 & 0.53$\pm$0.06 & 0.46$\pm$0.17 & \textbf{0.45}$\pm$0.18 & 10.50$\pm$0.23 \\ 
		Hyves & 6.18$\pm$0.06 & \textbf{0.30}$\pm$0.04 & 0.32$\pm$0.00 & 1.56$\pm$0.25 & 16.77$\pm$0.01 \\ 
		Cora & 16.60$\pm$0.06 & 1.43$\pm$0.11 & 0.95$\pm$0.16 & \textbf{0.20}$\pm$0.22 & 8.09$\pm$0.12 \\ 
		HepTh & 16.00$\pm$0.20 & 1.09$\pm$0.17 & 0.68$\pm$0.07 & \textbf{0.59}$\pm$0.25 & 15.73$\pm$0.15 \\ 
		HepPh & 25.39$\pm$0.11 & 0.34$\pm$0.11 & 0.64$\pm$0.11 & \textbf{0.28}$\pm$0.01 & 16.06$\pm$0.26 \\ 
		Topology & 1.70$\pm$0.07 & 0.91$\pm$0.11 & 0.31$\pm$0.24 & \textbf{0.22}$\pm$0.05 & 4.08$\pm$0.12 \\ 
		Gnutella & 23.43$\pm$0.24 & 2.89$\pm$0.10 & 2.67$\pm$0.08 & \textbf{2.36}$\pm$0.21 & 17.51$\pm$0.19 \\ 
		Caida & 9.59$\pm$0.13 & 0.68$\pm$0.01 & \textbf{0.34}$\pm$0.23 & 0.97$\pm$0.25 & 11.26$\pm$0.07 \\
		FF & 2.84$\pm$0.20 & 0.27$\pm$0.02 & 2.92$\pm$0.13 & \textbf{0.10}$\pm$0.06 & 5.63$\pm$0.09 \\
		SW & 88.03$\pm$0.00 & 4.79$\pm$0.01 & 9.27$\pm$0.12 & \textbf{3.15}$\pm$0.20 & 99.91$\pm$0.18 \\ 
		MM & 1.65$\pm$0.06 & 1.15$\pm$0.21 & 3.05$\pm$0.15 & \textbf{0.34}$\pm$0.02 & 9.86$\pm$0.01 \\ \hline
		Average & 17.37$\pm$0.12 & 1.11$\pm$0.08 & 1.58$\pm$0.12 & \textbf{0.81}$\pm$0.14 & 19.15$\pm$0.14 \\ \hline
	\end{tabular}
	\label{Tab_PL_RMSE}	
\end{figure}

The point statistics of path length in all the networks are shown in Figure~\ref{fig_PL_Stats}. The figure shows that HJ and FS perform poorly and most of their statistics fall out of the plotting area. We fixed the plotting area otherwise it would have been difficult to see the results visually because of the high statistics of HJ and FS. Other sampling algorithms extract good samples in most of the networks with a few exceptions, e.g., the small-world network could not be sampled well by any algorithm. We give the RMSE values with standard deviations in Table~\ref{Tab_PL_RMSE}. We find that HJ and FS samples produce big error values. Overall, the RMSE values are on the higher side when compared with that of the degree and clustering coefficient statistics. The possible reason is that the path length of a network is a complex metric and it seems that sampling algorithms need a more thoughtful design to sample path length. 

\begin{figure}[H]
	\centering
	\includegraphics[width = 120mm, height=150mm]{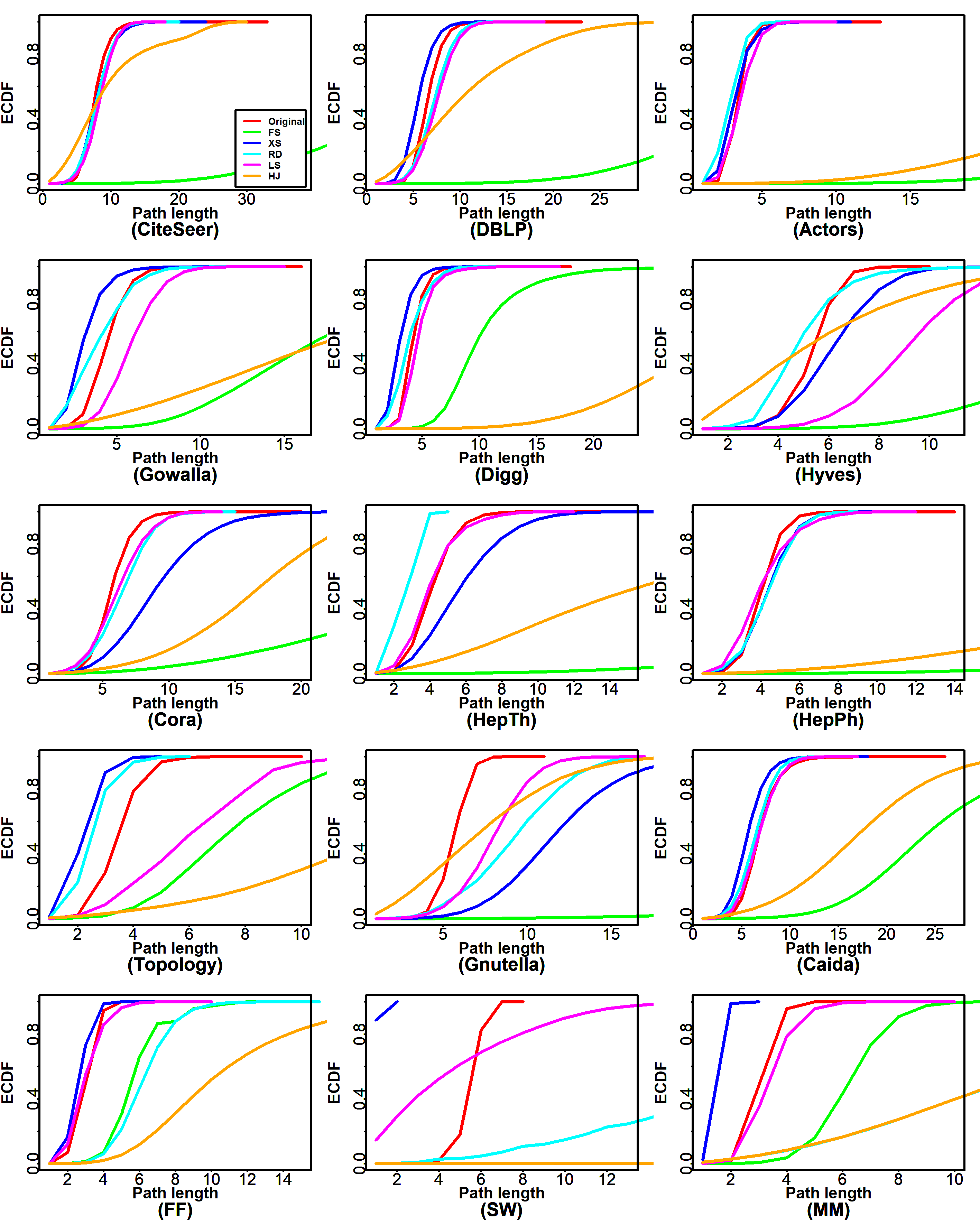}
	\caption{Path length distributions of all the networks at $\phi=0.02$. \textbf{(Best viewed in color.)} }
	\label{fig_PL_Dist}
	\vspace{0.75cm}
	\captionof{table}{Jensen Shannon distance and standard deviations for path length distributions. \textbf{Boldface} values are the best results.}
	\scalebox{0.95}{
		\begin{tabular}{|l|l|l|l|l|l|}
			\hline
			Datasets & \textbf{FS} & \textbf{XS} & \textbf{RD} & \textbf{LS} & \textbf{HJ} \\ \hline
			CiteSeer & 0.78$\pm$0.10 & \textbf{0.12}$\pm$0.04 & 0.13$\pm$0.04 & 0.16$\pm$0.06 & 0.47$\pm$0.02 \\ 
			DBLP & 0.79$\pm$0.09 & 0.27$\pm$0.01 & \textbf{0.17}$\pm$0.14 & 0.20$\pm$0.09 & 0.50$\pm$0.01 \\ 
			Actors & 0.81$\pm$0.07 & 0.18$\pm$0.06 & 0.29$\pm$0.00 & \textbf{0.14}$\pm$0.04 & 0.77$\pm$0.14 \\ 
			Gowalla & 0.74$\pm$0.08 & 0.42$\pm$0.08 & 0.34$\pm$0.05 & \textbf{0.33}$\pm$0.01 & 0.57$\pm$0.06 \\ 
			Digg & 0.73$\pm$0.14 & 0.43$\pm$0.12 & 0.30$\pm$0.10 & \textbf{0.13}$\pm$0.04 & 0.79$\pm$0.04 \\ 
			Hyves & 0.75$\pm$0.00 & \textbf{0.26}$\pm$0.04 & 0.29$\pm$0.07 & 0.60$\pm$0.00 & 0.57$\pm$0.03 \\ 
			Cora & 0.68$\pm$0.03 & 0.41$\pm$0.12 & 0.21$\pm$0.07 & \textbf{0.18}$\pm$0.07 & 0.63$\pm$0.09 \\ 
			HepTh & 0.76$\pm$0.01 & 0.27$\pm$0.14 & 0.46$\pm$0.14 & \textbf{0.11}$\pm$0.06 & 0.54$\pm$0.10 \\ 
			HepPh & 0.77$\pm$0.07 & \textbf{0.15}$\pm$0.03 & 0.17$\pm$0.03 & 0.21$\pm$0.01 & 0.69$\pm$0.13 \\ 
			Topology & 0.60$\pm$0.06 & 0.51$\pm$0.12 & \textbf{0.40}$\pm$0.11 & 0.44$\pm$0.11 & 0.64$\pm$0.03 \\ 
			Gnutella & 0.78$\pm$0.00 & 0.66$\pm$0.15 & 0.52$\pm$0.11 & \textbf{0.48}$\pm$0.08 & 0.55$\pm$0.14 \\ 
			Caida & 0.74$\pm$0.02 & 0.26$\pm$0.09 & 0.15$\pm$0.11 & \textbf{0.09}$\pm$0.11 & 0.57$\pm$0.06 \\ 
			FF & 0.68$\pm$0.13 & 0.18$\pm$0.14 & 0.69$\pm$0.02 & \textbf{0.14}$\pm$0.01 & 0.73$\pm$0.06 \\ 
			SW & 0.81$\pm$0.02 & 0.81$\pm$0.11 & 0.69$\pm$0.10 & \textbf{0.69}$\pm$0.06 & 0.79$\pm$0.02 \\ 
			MM & 0.73$\pm$0.09 & 0.78$\pm$0.07 & 0.64$\pm$0.12 & \textbf{0.17}$\pm$0.06 & 0.64$\pm$0.09 \\ \hline
			Average & 0.74$\pm$0.06 & 0.38$\pm$0.09 & 0.36$\pm$0.08 & \textbf{0.27}$\pm$0.05 & 0.63$\pm$0.07 \\ \hline
	\end{tabular} }
	\label{Tab_JS_PL}
	
\end{figure}

We show the distribution of path length in all the networks in Figure~\ref{fig_PL_Dist}. Again HJ and FS give poor results in almost all the datasets. Other methods perform well in the collaboration networks and also extract reasonable samples in the social and citation networks. In Table~\ref{Tab_JS_PL}, we give JS distance values with standard deviations. We find that, on average, LS samples give the least error among all the sampling methods and outperforms in ten networks. This experiment shows that the sampling algorithms fail to follow the path length distribution in a network and do not estimate the average path length very well.

\subsection{Global Clustering Coefficient}

\begin{figure}[H]
	\centering
	\includegraphics[width = 120mm, height=150mm]{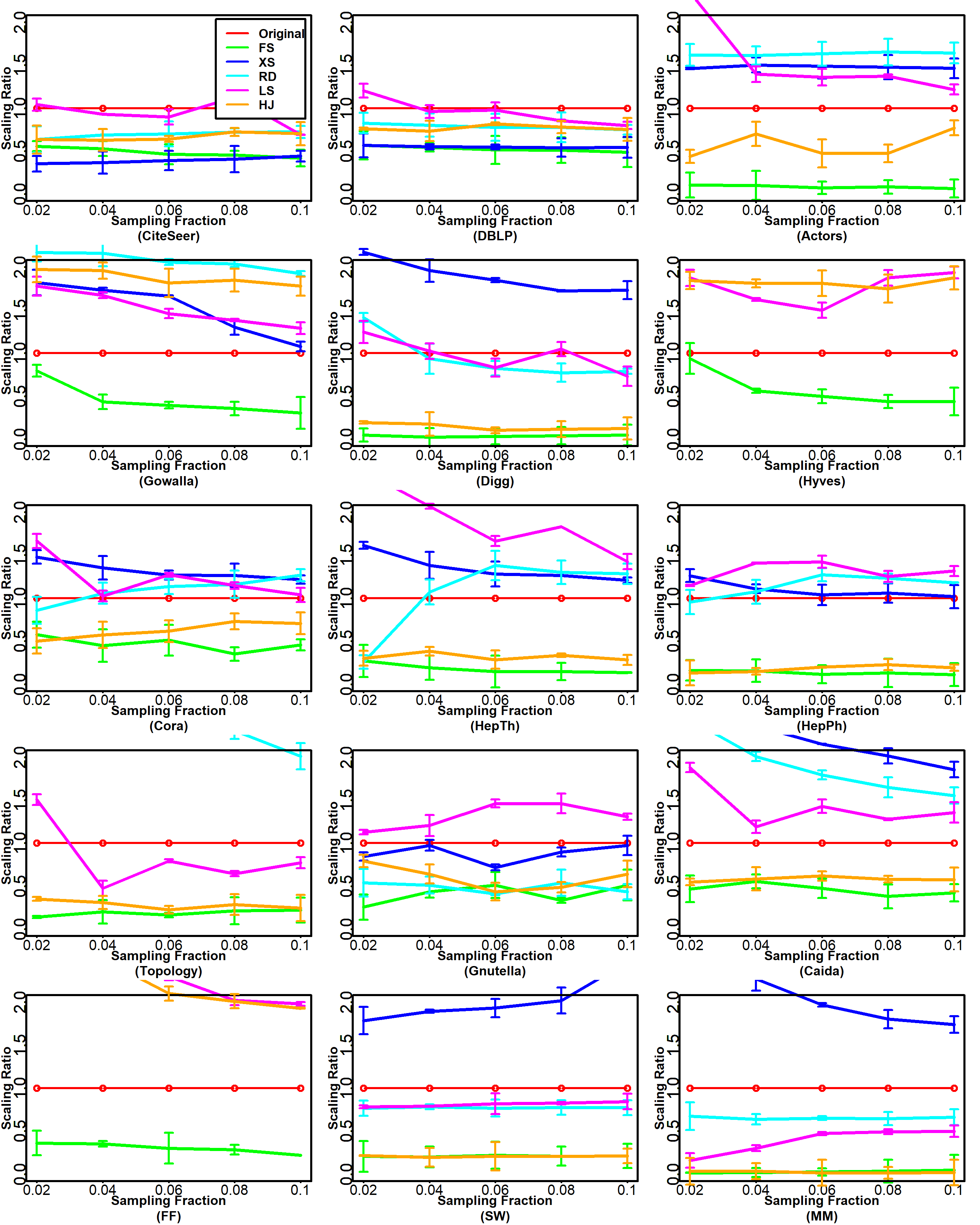}
	\caption{Point statistics of global clustering coefficient of all the networks with 95\% confidence intervals.}
	\label{fig_GCC_Stats}
	\vspace{0.75cm}
	\captionof{table}{RMSE values and standard deviations of point statistics of global clustering coefficient. \textbf{Boldface} values are the best results.}	
	\begin{tabular}{|l|l|l|l|l|l|}
		\hline
		Datasets & \textbf{FS} & \textbf{XS} & \textbf{RD} & \textbf{LS} & \textbf{HJ} \\ \hline
		CiteSeer & 0.23$\pm$0.14 & 0.27$\pm$0.08 & 0.13$\pm$0.17 & \textbf{0.05}$\pm$0.10 & 0.15$\pm$0.14 \\ 
		DBLP & 0.14$\pm$0.07 & 0.13$\pm$0.11 & 0.06$\pm$0.19 & \textbf{0.03}$\pm$0.04 & 0.07$\pm$0.01 \\ 
		Actors & 0.15$\pm$0.05 & 0.07$\pm$0.12 & 0.10$\pm$0.11 & 0.08$\pm$0.19 & \textbf{0.07}$\pm$0.08 \\ 
		Gowalla & 0.09$\pm$0.03 & 0.16$\pm$0.15 & 0.12$\pm$0.07 & \textbf{0.08}$\pm$0.04 & 0.11$\pm$0.01 \\ 
		Digg & 0.14$\pm$0.15 & 0.18$\pm$0.12 & 0.03$\pm$0.08 & \textbf{0.02}$\pm$0.21 & 0.14$\pm$0.17 \\ 
		Hyves & \textbf{0.07}$\pm$0.16 & 0.31$\pm$0.01 & 0.28$\pm$0.16 & 0.10$\pm$0.21 & 0.09$\pm$0.21 \\ 
		Cora & 0.06$\pm$0.07 & 0.03$\pm$0.09 & \textbf{0.01}$\pm$0.19 & 0.02$\pm$0.12 & 0.04$\pm$0.08 \\ 
		HepTh & 0.09$\pm$0.14 & 0.04$\pm$0.14 & \textbf{0.04}$\pm$0.03 & 0.14$\pm$0.12 & 0.08$\pm$0.16 \\ 
		HepPh & 0.12$\pm$0.17 & \textbf{0.01}$\pm$0.14 & 0.02$\pm$0.10 & 0.04$\pm$0.02 & 0.11$\pm$0.02 \\ 
		Topology & 0.07$\pm$0.05 & 0.28$\pm$0.16 & 0.22$\pm$0.16 & \textbf{0.04}$\pm$0.05 & 0.06$\pm$0.12 \\ 
		Gnutella & 0.06$\pm$0.08 & 0.02$\pm$0.08 & \textbf{0.04}$\pm$0.03 & 0.04$\pm$0.05 & 0.05$\pm$0.14 \\ 
		Caida & \textbf{0.03}$\pm$0.11 & 0.17$\pm$0.04 & 0.15$\pm$0.10 & 0.04$\pm$0.21 & 0.04$\pm$0.18 \\ 
		FF & \textbf{0.07}$\pm$0.15 & 0.21$\pm$0.12 & 0.23$\pm$0.04 & 0.19$\pm$0.15 & 0.20$\pm$0.17 \\ 
		SW & 0.11$\pm$0.06 & 0.23$\pm$0.13 & 0.08$\pm$0.02 & \textbf{0.06}$\pm$0.01 & 0.12$\pm$0.20 \\ 
		MM & 0.07$\pm$0.07 & 0.19$\pm$0.04 & \textbf{0.04}$\pm$0.15 & 0.06$\pm$0.06 & 0.07$\pm$0.15 \\ \hline
		Average & 0.10$\pm$0.10 & 0.15$\pm$0.10 & 0.11$\pm$0.11 & \textbf{0.06}$\pm$0.10 & 0.09$\pm$0.12 \\ \hline
	\end{tabular}
	\label{Tab_GCC_RMSE}	
\end{figure}

We present the results of point statistics of the global clustering coefficient of all the networks in Figure~\ref{fig_GCC_Stats} with 95\% confidence intervals. We see big variations in point statistics in all the networks in the samples. It seems there is no clear winner as one algorithm performs better in one dataset and worse in the other. For example, LS performs the best in the CiteSeer network but the worst in the HepTh network. Overall, LS works better on the collaboration and social networks whereas RD extracts better samples from the citation networks. It should be noted that LS and RD also performed well in matching the value of the average clustering coefficient in these networks. FS, XS, and HJ also extract some good samples. The RMSE values and standard deviations of these point statistics are presented in Table~\ref{Tab_GCC_RMSE}. LS and RD give the minimum error in six and four networks respectively. On average, LS outperforms other methods.  Although HJ does not outperform in individual networks, it stands second to LS on average. 

\subsection{Assortativity}
\begin{figure}[H]
	\centering
	\includegraphics[width = 120mm, height=150mm]{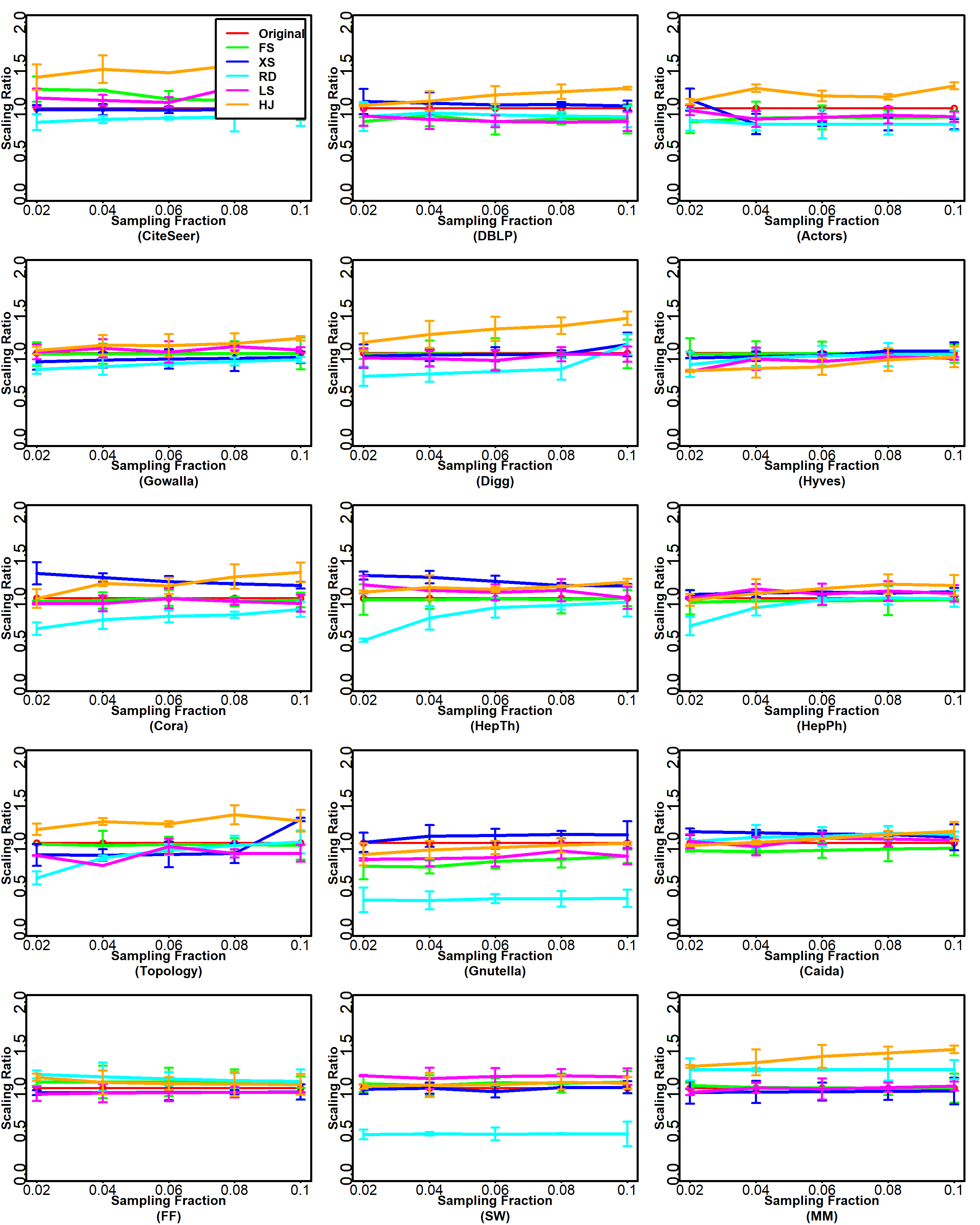}
	\caption{Point statistics of assortativity of all the networks with 95\% confidence intervals.}
	\label{fig_Assort_Stats}
	\vspace{0.75cm}
	\captionof{table}{RMSE values and standard deviations of point statistics of assortativity. \textbf{Boldface} values are the best results.}	
		\begin{tabular}{|l|l|l|l|l|l|}
		\hline
		Datasets & \textbf{FS} & \textbf{XS} & \textbf{RD} & \textbf{LS} & \textbf{HJ} \\ \hline
		CiteSeer & 0.13$\pm$0.24 & \textbf{0.01}$\pm$0.24 & 0.13$\pm$0.17 & 0.11$\pm$0.13 & 0.48$\pm$0.15 \\ 
		DBLP & 0.16$\pm$0.10 & \textbf{0.06}$\pm$0.17 & 0.10$\pm$0.12 & 0.17$\pm$0.01 & 0.17$\pm$0.14 \\ 
		Actors & 0.14$\pm$0.05 & 0.20$\pm$0.01 & 0.21$\pm$0.25 & \textbf{0.10}$\pm$0.01 & 0.20$\pm$0.15 \\ 
		Gowalla & \textbf{0.00}$\pm$0.15 & 0.06$\pm$0.05 & 0.12$\pm$0.10 & 0.03$\pm$0.15 & 0.09$\pm$0.05 \\ 
		Digg & \textbf{0.00}$\pm$0.24 & 0.03$\pm$0.19 & 0.18$\pm$0.01 & 0.04$\pm$0.08 & 0.24$\pm$0.07 \\ 
		Hyves & \textbf{0.00}$\pm$0.20 & 0.03$\pm$0.13 & 0.04$\pm$0.12 & 0.09$\pm$0.13 & 0.12$\pm$0.02 \\ 
		Cora & \textbf{0.02}$\pm$0.23 & 0.19$\pm$0.17 & 0.21$\pm$0.17 & 0.04$\pm$0.02 & 0.16$\pm$0.05 \\ 
		HepTh & \textbf{0.01}$\pm$0.13 & 0.19$\pm$0.08 & 0.18$\pm$0.02 & 0.07$\pm$0.11 & 0.12$\pm$0.14 \\ 
		HepPh & \textbf{0.02}$\pm$0.09 & 0.06$\pm$0.06 & 0.09$\pm$0.06 & 0.05$\pm$0.05 & 0.09$\pm$0.13 \\ 
		Topology & \textbf{0.01}$\pm$0.15 & 0.12$\pm$0.03 & 0.10$\pm$0.10 & 0.10$\pm$0.01 & 0.19$\pm$0.06 \\ 
		Gnutella & 0.19$\pm$0.10 & 0.06$\pm$0.17 & 0.59$\pm$0.21 & 0.14$\pm$0.15 & \textbf{0.05}$\pm$0.03 \\ 
		Caida & 0.08$\pm$0.07 & 0.10$\pm$0.17 & 0.07$\pm$0.16 & \textbf{0.04}$\pm$0.05 & 0.06$\pm$0.08 \\
		FF & 0.06$\pm$0.01 & \textbf{0.04}$\pm$0.20 & 0.10$\pm$0.05 & 0.05$\pm$0.12 & 0.06$\pm$0.04 \\ 
		SW & 0.05$\pm$0.18 & \textbf{0.01}$\pm$0.08 & 0.53$\pm$0.17 & 0.13$\pm$0.07 & 0.04$\pm$0.07 \\ 
		MM & \textbf{0.00}$\pm$0.01 & 0.03$\pm$0.17 & 0.17$\pm$0.16 & 0.01$\pm$0.01 & 0.29$\pm$0.02 \\ \hline
		Average & \textbf{0.06}$\pm$0.13 & 0.08$\pm$0.13 & 0.19$\pm$0.12 & 0.08$\pm$0.07 & 0.16$\pm$0.08 \\ \hline
	\end{tabular}
	\label{Tab_Assort_RMSE}	
\end{figure}

We show the point statistics of assortativity of all the networks in Figure~\ref{fig_Assort_Stats} with 95\% confidence intervals. It should be noted that assortativity lies in the range [-1, +1], however, to calculate the point statistics we converted it to the range [0, 2] for all the sample and original graphs. Interestingly, all the sampling methods show good results. It seems that these algorithms can extract good samples while preserving degree mixing patterns of nodes. In particular, FS seems to extract better samples than the other methods from the social and citation networks whereas XS and LS also perform better in few networks. Table~\ref{Tab_Assort_RMSE} gives the RMSE values of all the sampling methods along with standard deviations. We find that, on average, FS samples give the smallest error values whereas error values of XS and LS are also reasonably small.

\subsection{Modularity}
\begin{figure}[H]
	\centering
	\includegraphics[width = 120mm, height=150mm]{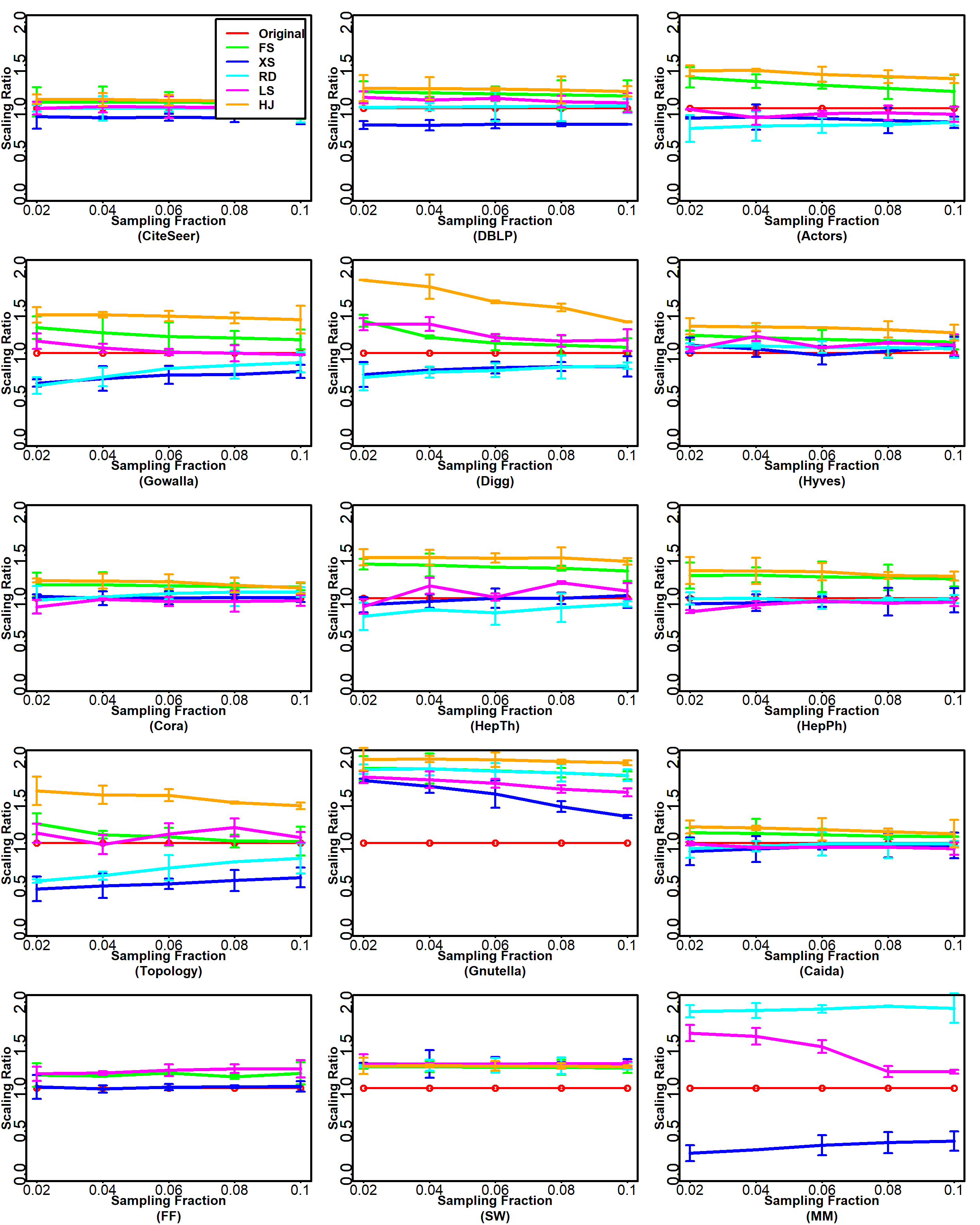}
	\caption{Point statistics of modularity of all the networks with 95\% confidence intervals.}
	\label{fig_Mod_Stats}
	\vspace{0.75cm}
	\captionof{table}{RMSE values and standard deviations of point statistics of modularity. \textbf{Boldface} values are the best results.}	
	\begin{tabular}{|l|l|l|l|l|l|}
		\hline
		Datasets & \textbf{FS} & \textbf{XS} & \textbf{RD} & \textbf{LS} & \textbf{HJ} \\ \hline
		CiteSeer & 0.06$\pm$0.09 & 0.09$\pm$0.15 & \textbf{0.00}$\pm$0.03 & 0.00$\pm$0.03 & 0.08$\pm$0.01 \\ 
		DBLP & 0.13$\pm$0.18 & 0.15$\pm$0.07 & \textbf{0.01}$\pm$0.03 & 0.07$\pm$0.04 & 0.17$\pm$0.03 \\ 
		Actors & 0.18$\pm$0.04 & 0.08$\pm$0.10 & 0.13$\pm$0.11 & \textbf{0.04}$\pm$0.12 & 0.27$\pm$0.08 \\ 
		Gowalla & 0.14$\pm$0.17 & 0.18$\pm$0.13 & 0.14$\pm$0.13 & \textbf{0.03}$\pm$0.16 & 0.29$\pm$0.01 \\ 
		Digg & \textbf{0.08}$\pm$0.11 & 0.09$\pm$0.07 & 0.10$\pm$0.18 & 0.12$\pm$0.16 & 0.33$\pm$0.08 \\ 
		Hyves & 0.12$\pm$0.10 & \textbf{0.04}$\pm$0.07 & 0.05$\pm$0.04 & 0.08$\pm$0.15 & 0.21$\pm$0.00 \\ 
		Cora & 0.11$\pm$0.07 & \textbf{0.00}$\pm$0.17 & 0.03$\pm$0.17 & 0.03$\pm$0.10 & 0.13$\pm$0.16 \\ 
		HepTh & 0.23$\pm$0.17 & \textbf{0.01}$\pm$0.06 & 0.08$\pm$0.05 & 0.06$\pm$0.04 & 0.29$\pm$0.01 \\ 
		HepPh & 0.18$\pm$0.08 & 0.03$\pm$0.16 & \textbf{0.00}$\pm$0.12 & 0.05$\pm$0.07 & 0.21$\pm$0.11 \\ 
		Topology & \textbf{0.05}$\pm$0.12 & 0.28$\pm$0.09 & 0.18$\pm$0.11 & 0.05$\pm$0.01 & 0.31$\pm$0.05 \\ 
		Gnutella & 0.41$\pm$0.15 & \textbf{0.26}$\pm$0.08 & 0.41$\pm$0.14 & 0.34$\pm$0.12 & 0.47$\pm$0.12 \\ 
		Caida & 0.08$\pm$0.07 & 0.04$\pm$0.14 & \textbf{0.01}$\pm$0.05 & 0.03$\pm$0.08 & 0.12$\pm$0.07 \\ 
		FF & 0.04$\pm$0.09 & \textbf{0.00}$\pm$0.10 & 0.49$\pm$0.11 & 0.05$\pm$0.07 & 0.54$\pm$0.14 \\ 
		SW & \textbf{0.19}$\pm$0.11 & 0.21$\pm$0.14 & 0.20$\pm$0.11 & 0.21$\pm$0.15 & 0.19$\pm$0.10 \\ 
		MM & 0.34$\pm$0.10 & 0.10$\pm$0.10 & 0.14$\pm$0.10 & \textbf{0.06}$\pm$0.10 & 0.77$\pm$0.00 \\ \hline
		Average & 0.16$\pm$0.11 & 0.10$\pm$0.11 & 0.13$\pm$0.10 & \textbf{0.08}$\pm$0.09 & 0.29$\pm$0.07 \\ \hline
	\end{tabular}
	\label{Tab_Mod_RMSE}	
\end{figure}

In Figure~\ref{fig_Mod_Stats}, we present the point statistics of modularity of all the networks with 95\% confidence intervals. Each method, except HJ, preserves modularity in few networks showing that these methods can extract community structure from the graph being sampled. There is one exception of the technological network Gnutella in which all the methods fail to extract good samples, especially at small sampling fractions. FS and HJ perform poorly in the synthetic network of MM. We give the RMSE values and standard deviations in Table~\ref{Tab_Mod_RMSE}. The table shows that XS and RD give the smallest error in five and four networks respectively while FS and LS outperform in three networks each. On average, LS samples give the minimum error in this metric. 

\subsection{Summary and Discussion}
We summarize the average values of Root Mean Square Error (RMSE) for point statistics and that of Jensen-Shannon Distance (JSD) for distributions of all the sampling methods in Table~\ref{Tab_Summary}. The table shows that the properties of the samples extracted by LS better match with the properties of the original networks. A possible explanation is that 1) LS prefers to sample the nodes from the neighborhood of already sampled nodes. The inclusion of neighboring nodes helps LS in extracting the structure around a node that results in maintaining the overall structure of the given graph. 2) The induction step of LS augments all the edges among the sampled nodes that directly helps in estimating properties such as clustering coefficient. The RD method prefers high degree nodes and extracts samples with a better degree estimation than other methods. The m-dimensional random walks in FS seem to explore the nodes of varying degrees and hence it can better match the degree mixing pattern of nodes. We believe that graph sampling is a complex process and given the fact that every real-world graph has a different set of property values, it seems very difficult to design a single sampling algorithm that can work for all real-world networks. 

\begin{table}
	\centering
	\caption{Summary of the results: The average values of RMSE and JSD of all the sampling algorithms. \textbf{Boldface} values are the best results.}
	\begin{tabular}{|l|l|l|l|l|l|l|}
		\hline
		Metric &Graph Property & FS & XS & RD & LS  & HJ \\ \hline
		\multirow{6}{*}{RMSE}
		&Degree & 13.06 & 20.25 & \textbf{7.74} & 10.52 & 14.16 \\ 
		&Clustering Coefficient & 0.30 & 0.14 & 0.10 & \textbf{0.04} & 0.30 \\ 
		&Path length & 17.37 & 1.11 & 1.58 & \textbf{0.81} & 19.15 \\ 
		&Global Clustering Coefficient & 0.10 & 0.15 & 0.11 & \textbf{0.06} & 0.09 \\ 
		&Assortativity & \textbf{0.06} & 0.08 & 0.19 & 0.08 & 0.16 \\ 
		&Modularity & 0.16 & 0.10 & 0.13 & \textbf{0.08} & 0.29 \\ \hline
		\multirow{3}{*}{JSD}
		&Degree & 0.47 & 0.27 & 0.32 & \textbf{0.24} & 0.32 \\
		&Clustering Coefficient & 0.48 & 0.35 & 0.33 & \textbf{0.22} & 0.49 \\ 
		&Path length & 0.74 & 0.38 & 0.36 & \textbf{0.27} & 0.63 \\ \hline
	\end{tabular}
     \label{Tab_Summary}
\end{table}

\section{Related Work}
Graph sampling is used to obtain a representative subgraph from a large graph. A very naive approach is to select nodes uniformly at random from the original graph and then induce the sample graph over this node-set. A similar approach is to sample the edges uniformly at random and then add the nodes at the ends of those edges to the node-set. A fundamental problem in these random sampling methods is that real-world large graphs are not available as a whole and it becomes nearly infeasible to select nodes or edges uniformly at random. In traversal based sampling we traverse a small portion of a graph, e.g., Breadth First Sampling \citep{BFS_2}, Random First Sampling \citep{Metric}, Snowball Sampling \citep{statistical} and different variations of Random Walk \citep{Estimating, RW_index,RW_p2p,RW_p2p_2,RW_RD}. More recently, the authors \citep{GS} apply a traversal based sampling that consumes the local information of nodes, combined with the estimated values of a set of properties, to guide the sampling process and extract tiny samples that preserve the properties of the graph. All these approaches have their own pros and cons. For example, Breadth First Sampling (BFS) is known to overestimate samples in terms of degree because BFS samples are biased towards high degree nodes. The closest to our work is the one presented in \citep{Wang} and provides a good understanding of how sampling works in big graphs. The authors analyze several graph sampling algorithms and evaluate their performance on some widely recognized graph properties on directed graphs using large-scale social network datasets. However, the work considers a very small set of properties of a graph compared to our work.

\section{Conclusion}
In this work, we conduct a comprehensive empirical study to characterize several graph sampling algorithms. We test the ability of these sampling methods in maintaining the properties of the original graphs. We characterize these methods on five types of graphs and test their ability to match six properties of the original graphs. We find that the algorithms that explore the neighborhood of a sampled node on priority bases perform better than other methods in maintaining the structure of the original graph in their samples. We also find that preferring high degree nodes during the sampling process also benefits the sampling methods. However, we realize that extracting the structure and maintaining the properties of a graph in a tiny sample is a difficult task and needs a very thoughtful sampling process.


%
%

\bibliographystyle{spbasic}      
\bibliography{OverSampling_Bib}   

\end{document}